\documentclass[epj]{svjour}

\usepackage{graphicx,epsfig,subfigure}
\usepackage{amsmath,amssymb,cite}

\begin{document}

\title{The effect of confinement on stochastic resonance in continuous bistable systems}
\author{Shubhashis Rana, Sourabh Lahiri and A. M. Jayannavar}
\institute{Institute of Physics, Sachivalaya Marg, Bhubaneswar-751005, India}

\date{}


\newcommand{\nwc}{\newcommand}
\nwc{\la}{\langle}
\nwc{\ra}{\rangle}
\nwc{\lw}{\linewidth}
\nwc{\nn}{\nonumber}
\nwc{\Ra}{\Rightarrow}

\nwc{\pd}[2]{\frac{\partial #1}{\partial #2}}
\nwc{\zprl}[3]{Phys. Rev. Lett. ~{\bf #1},~#2~(#3)}
\nwc{\zpre}[3]{Phys. Rev. E ~{\bf #1},~#2~(#3)}
\nwc{\zpra}[3]{Phys. Rev. A ~{\bf #1},~#2~(#3)}
\nwc{\zjsm}[3]{J. Stat. Mech. ~{\bf #1},~#2~(#3)}
\nwc{\zepjb}[3]{Eur. Phys. J. B ~{\bf #1},~#2~(#3)}
\nwc{\zrmp}[3]{Rev. Mod. Phys. ~{\bf #1},~#2~(#3)}
\nwc{\zepl}[3]{Europhys. Lett. ~{\bf #1},~#2~(#3)}
\nwc{\zjsp}[3]{J. Stat. Phys. ~{\bf #1},~#2~(#3)}
\nwc{\zptps}[3]{Prog. Theor. Phys. Suppl. ~{\bf #1},~#2~(#3)}
\nwc{\zpt}[3]{Physics Today ~{\bf #1},~#2~(#3)}
\nwc{\zap}[3]{Adv. Phys. ~{\bf #1},~#2~(#3)}
\nwc{\zjpcm}[3]{J. Phys. Condens. Matter ~{\bf #1},~#2~(#3)}
\nwc{\zjpa}[3]{J. Phys. A ~{\bf #1},~#2~(#3)}
\nwc{\zpjp}[3]{Pram. J. Phys. ~{\bf #1},~#2~(#3)}
\nwc{\zpa}[3]{Physica A ~{\bf #1},~#2~(#3)}

\abstract{
     Using the input energy per cycle as a quantifier of stochastic resonance (SR), we show that SR is observed in superharmonic (hard) potentials. However, it is not observed in subharmonic (soft) potentials, even though the potential is bistable. These results are consistent with recent observations based on amplitude of average position as a quantifier. In both soft and hard potentials, we observe resonance phenomenon as a function of the driving frequency. The nature of probability distributions of average work are qualitatively different for soft and hard potentials.
}
 
\authorrunning{Subhashis Rana et. al.}
\titlerunning{Effect of confinement on SR}
\maketitle{}

\vspace{1cm}

\section{Introduction}

Stochastic resonance is an exclusively nonlinear phenomenon where the combined effect of the noise and the nonlinearity gives rise to an enhanced response of the system at a particular frequency of an external periodic drive. It has been found to be of fundamental importance not only in physics \cite{han98,mar09,wie95,wei95a} but also in biological systems, from the mechanoreceptor cells in crayfish to the functioning of sensory neurons in humans. This is in sharp contrast to the general trend of a noise to cause the effect of a signal to fade. A typical model used to study this behaviour consists of a bistable potential in which a Brownian particle is present. The particle is in contact with a thermal bath of temperature $T$. This system is driven by a periodic drive with a given frequency,  $f(t)=A\sin\omega t$. Now, the initial system without the drive has an intrinsic escape rate of going from one minimum of the potential ($V(x)$) to the other. This is given by the \emph{Kramers escape rate} \cite{ris,cha43}:
\begin{equation}
r_K = C e^{-\beta\Delta V},
\end{equation}
where $\beta\equiv 1/k_BT$, $k_B$ being the Boltzmann Constant, $C$ is a constant that depends on the system parameters, and $\Delta V$ is the barrier height (height difference between the minimum and the maximum of $V(x)$). The escape time will then be given by the inverse of the escape rate: $\tau_K = r_K^{-1}$. As the external periodic drive is switched on, in general its time period $\tau_\omega$ will not be in synchronization with the escape time of the particle over the barrier. 
However, if the noise strength or temperature is varied, at a certain value of temperature, $\tau_\omega$ will be exactly equal to $2\tau_K$. Now we will have proper synchronization of the dynamics: when the right well becomes deeper compared to the left well, the particle hops into the right well with a high probability, and vice versa. 
Under this condition,  the system absorbs  maximum energy from the drive. This phenomenon is termed as \emph{stochastic resonance} (SR). Various quantifiers of SR have been proposed in literature \cite{han98}: signal-to-noise ratio (SNR), hysteresis loop area (HLA), spectral power amplification (SPA), position amplitude of the particle ($\bar x$), phase lag of the response with the external drive ($\bar\phi$), etc.  
In \cite{evs05}, a relation has been derived between SNR, HLA and SPA which is given by:
\begin{equation}
\mbox{SNR} \times \mbox{HLA} = -\frac{\pi^2 A^4 \omega}{4k_BT}\mbox{SPA}
\end{equation}
The above relation shows that some of the above quantifiers are related to each other.
We observe that a similar relation is present that connects the quantifiers mean thermodynamic work, $\bar x$ and $\bar\phi$, and as a result the three cannot act as independent quantifiers of SR.

In the present work we use the mean input energy per drive period as a quantifier for SR \cite{iwa01,dan03,mam08,roy07}. As a function of noise strength, we observe suppression of SR in soft potentials and distinct peak signifying SR in hard potentials. 

 Since it has been established that SR is a bonafide resonance 
it should show maximum in the quantifiers as a function of drive frequency 
as well. Interestingly, whereas the mean input energy per period, $\la W\ra$, shows 
peaking behaviour, the average amplitude, $\overline{x}$ does not. As opposed to the behaviour of $\la W\ra$ as a function of noise strength, for soft and hard potentials, its behaviour with frequency is same for both kinds of potentials. In particular, $\la W\ra$ shows peaking behaviour and $\bar x$ decreases monotonically as a function of frequency for both hard and soft potentials.
 We further investigate the probability distributions of work for various temperatures \cite{mam08,roy07}, for both hard and soft potentials, and point out the qualitative differences between the nature of these distributions.

\section{The system}

\begin{figure}
\includegraphics[width=6cm]{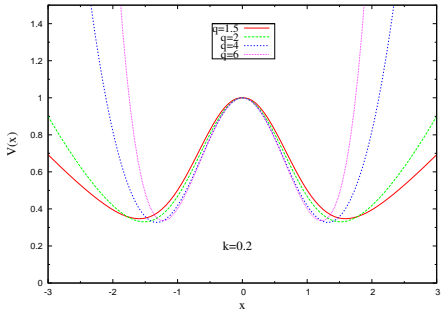}
\caption{The shapes of the potential for different values of $q$, with $k=0.2$.}
\label{potfig}
\end{figure}

We consider a Brownian particle described by the overdamped Langevin equation:
\begin{equation}
\gamma\dot x = -V'(x)+f(t)+\xi(t),
\end{equation}
where $V'(x)=\pd{V(x)}{x}$, $\xi(t)$ is Gaussian distributed white noise having the properties $\la\xi(t)\ra=0$ and $\la\xi(t)\xi(t')\ra=2\gamma k_BT\delta(t-t')$, $\gamma$ being the viscous drag in the medium. The strength of the noise is described by the thermal energy $k_BT$. The expression for the bistable potential is given by \cite{mar10}
\begin{equation}
V(x) = e^{-x^2} + k\frac{|x|^q}{q},
\label{pot}
\end{equation}
where the parameter $k$ has been set equal to 0.2 throughout the manuscript, which sets the barrier height at approximately $\Delta V = 0.67$ (all variables are dimensionless). The parameter $q$ is used to modulate the steepness of the walls of the potential. In other words, as $q$ is increased, the slope of the potential wall  increases as shown in figure \ref{potfig}, so that the particle is more confined in-between the two minima. The external drive is given by $f(t)=A\sin\omega t$.

If we plot the amplitude of mean position as a function of temperature, i.e., $\bar x(T)$ vs $T$, then the resonance gets suppressed when $q\le 2$. Analytically, this follows from the following approximate expression for $\la x\ra$ in the nonequilibrium steady state (where the intrawell dynamics has been ignored) \cite{han98}:
\begin{equation}
\la x(t)\ra = \bar x \sin(\omega t-\bar\phi)
\label{response}
\end{equation}
where 
\begin{equation}
\bar x(T) = \frac{A\la x^2\ra_0}{T}\frac{2r_K}{\sqrt{4r_K^2+\omega^2}}.
\label{xTavg}
\end{equation}
\begin{equation}
\bar\phi(T) = \tan^{-1}\left(\frac{\omega}{2r_K}\right)
\label{phi}
\end{equation}
Here, the angular brackets $\la\cdots\ra$ represent ensemble averaging over a large number of phase space trajectories.
$\la x^2\ra_0$ is the variance of the position of the particle in {\it absence} of any drive, i.e., subjected to the unperturbed potential.
 $r_K$ is the Kramers escape rate whose expression is given by \cite{ris,cha43}
\begin{equation}
r_K = \left(\frac{\gamma\sqrt{V''(x_m).|V''(0)|}}{2\pi}\right)e^{-\beta\Delta V},
\label{r}
\end{equation}
where $\pm x_{m}$ and $0$ are the positions of the minima and of the maximum of the potential,  $\Delta V$ is the barrier height, and $V''$ is double derivative of the potential function with respect to $x$.

It can then be shown from (\ref{xTavg}) that as $T\to\infty$, the behaviour of the amplitude of mean position is given by \cite{mar10}
\begin{equation}
\lim_{T\to\infty}\bar x(T) \sim T^{2/q-1}.
\label{xTamp}
\end{equation}
Now it can easily be seen that if $q>2$, then $\bar x(T)$ goes to zero for large $T$. Of course, as $T\to 0$, $\bar x\to 0$ as well, because the particle hardly deviates from its equilibrium position (for details refer to \cite{mar10}). This means that there must be a maximum in-between these two limits - a signal for resonance. However, if $q<2$, the particle travels large distances away from the minima, so that the $\bar x$ grows monotonically with temperature, and stochastic resonance is not observed. The case $q=2$ is the marginal case.

Following stochastic energetics \cite{sek97}, the thermodynamic work done on the system is given by
\begin{equation}
W = \int_0^\tau \pd{V(x,t)}{t}dt = -\int_0^\tau x(t)\frac{df(t)}{dt}dt,
\label{Wdef}
\end{equation}
where $V(x,t) = V(x)-xf(t)$.

The average work done over time $\tau$ is 
\[
\la W\ra = -\int_0^\tau \la x(t)\ra\frac{df(t)}{dt}dt.
\label{Wavgdef}
\]
Now, using the approximate expression for $\la x(t)\ra$ (eq. (\ref{response})), we get
\begin{eqnarray}
\la W\ra &=& -\int_0^\tau \la x(t)\ra \dot f(t)dt \nn\\
&=& -A\omega\int_0^\tau \bar x \sin(\omega t-\bar\phi)\cos\omega t dt \nn\\
&=& A~\frac{2\pi}{\tau}~\bar x~\sin\bar\phi~\frac{\tau}{2} \nn\\
&=& A\pi\bar x\sin\bar\phi.
\label{W1}
\end{eqnarray}
Thus, we have arrived at a relation that connects three of the   proposed quantifiers of SR: $\la W\ra$, $\bar x$ and $\bar\phi$.
Plugging in the expressions for $\bar x(T)$ (eq. (\ref{xTavg})) and $\bar\phi(T)$ (eq.(\ref{phi})), we find
\begin{equation}
\la W\ra = \frac{2\pi A^2\la x^2\ra_0~\omega r_K}{T(4r_K^2+\omega^2)}.
\label{W}
\end{equation}

Eq. (\ref{W}) predicts SR as a function of temperature. Later on we use this to analyze our numerical results.

\section{Results and discussions}

\begin{figure}[h]
\includegraphics[width=7cm]{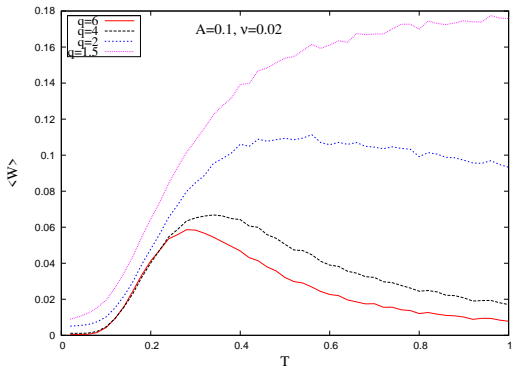}
\caption{Average work as a function of temperature for $A=0.1$ and $\nu=0.02$ for different values of $q$. SR is observed for all values of $q$ excepting $q=1.5$.}
\label{SRT}
\end{figure}

The plots of average thermodynamic work done on the particle by the drive per period $\tau_\omega$ have been shown, for different values of the parameter $q$ in figure \ref{SRT}. 
These plots have been obtained numerically by using the Heun's method \cite{man00}. We have ignored the initial transients and have evaluated the work over many cycles ($\sim 10^5$) using a single long trajectory of the particle.
We find that the plots qualitatively show the same features as shown by the position amplitude with temperature \cite{mar10}. The parameters used have been given in the figure captions.
  At very low temperature, the particle can see only a single well, the barrier height being too large for it to cross. Thus, only intrawell dynamics is dominant under this condition and as a result the work done on the particle is very small and goes to zero as $T\to 0$. At the other extreme, $T\to\infty$, the barrier becomes negligible compared to the thermal energy of the particle. Now the random motion of the particle becomes so large that the synchronization gets washed away.   This happens only for strong confining potential with $q>2$. Thus for the hard potentials, we get a clear resonance peak. Moreover, our numerical result shows that the temperature at which SR peak occurs is consistent with the condition $\tau_\omega = 2\tau_K$.

However, for values of $q\le 2$, the particle travels far from the left(right) of left(right) minimum and takes a long time to return. So it is expected that the distribution of passage time above the barrier will be very broad, so that the mean passage time ceases to be a good variable, being dominated by a large dispersion. As a result, the synchronization condition of escape rate with drive period is never satisfied. Thus these plots do not show the characteristic maxima of SR. 

\begin{figure}[h]
\includegraphics[width=7cm]{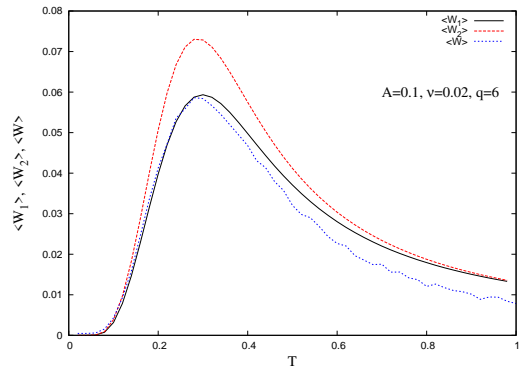}
\caption{Comparison of temperature dependence of $\la W\ra$ obtained numerically with the analytical results. $\la W_1\ra$ (the smooth solid line) is the expression obtained by taking into account the temperature dependence of the variance in position of the particle, $\la x^2\ra_0$. $\la W_2\ra$ (the smooth dashed line) is the expression for average work obtained with the variance replaced by $x_m^2$. The curve labelled $\la W\ra$ is the numerically generated one.}
\label{ana_compare}
\end{figure}

In figure \ref{ana_compare}, the numerically obtained plot for $\la W\ra$ vs $T$ has been compared for the superharmonic potential with $q=6$ with the following two analytical expressions for average work in steady state that are commonly used in the literature \cite{mar10}:

\begin{subequations}
\begin{equation}
\la W_1\ra = \dfrac{2\pi A^2\la x^2\ra_0}{T}~\dfrac{\omega r_K}{4r_K^2+\omega^2}.
\label{W1}
\end{equation}

\begin{equation}
 \la W_2\ra = \dfrac{2\pi A^2x_m^2}{T}~\dfrac{\omega r_K}{4r_K^2+\omega^2}.
\label{W2}
\end{equation}

\end{subequations}

In the expression for $\la W_2\ra$, we have replaced $\la x^2\ra_0$ by $x_m^2$, where $x_m$ is the position of the minimum of the potential.  However, from the figure we find that $\la W_1\ra$ matches the simulated curve reasonably well, whereas $\la W_2\ra$ deviates by a larger extent. This can be understood as follows \cite{mar10}. In the derivation of the expression for $\la W_2\ra$, one assumes that we are dealing with a {\it strictly} two-state system, where the particle's position distribution is the summation of two delta functions. $x_m^2$ then becomes the variance of the distribution having two $\delta$-functions equidistant from the origin: $\la x_m-0\ra^2 = x_m^2$. Evidently, at any finite temperature the above distribution will be incorrect, owing to the softness of the double well potential. Thus, we need to incorporate into our expression the temperature dependence of the variance in particle position, which has been done in deriving the expression for $\la W_1\ra$.

\begin{figure}[h]
\includegraphics[width=7cm]{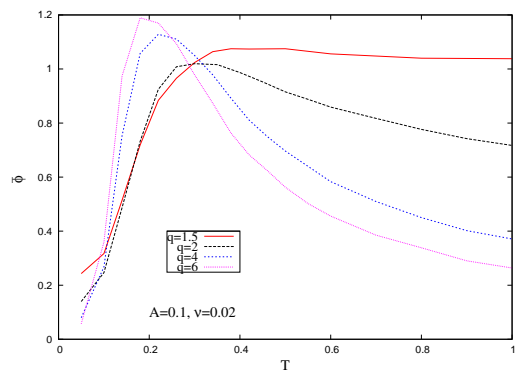}
\caption{Phase difference $\bar\phi$ between drive and response, as a function of bath temperature. The peak in the curve becomes less distinct as the value of $q$ is lowered.}
\label{phi_T}
\end{figure}

Previously the phase lag $\bar \phi$ of the response with drive (eq. (\ref{response})) has been used to detect stochastic resonance \cite{dyk92}. The variation of $\bar \phi$ with temperature has been shown in figure \ref{phi_T}. We find that the curves for $q=2,$ 4 and 6 show prominent maxima, whereas the curve for $q=1.5$ monotonically increases from zero and then saturates to an upper limit. 
Systems exhibiting SR show a peak in the phase lag $\bar\phi$ as a function of noise strength. However, the optimum value of noise intensity at which peak occurs does not coincide with SR peak for different quantifiers. The bell-shaped dependence reflects the competition between hopping and intrawell dynamics \cite{han98}.
However, for $q=1.5$, we do not see a bell-shaped curve, implying no clear-cut time scale separation between hopping and intrawell motion.

\begin{figure}[h]
\includegraphics[width=7cm]{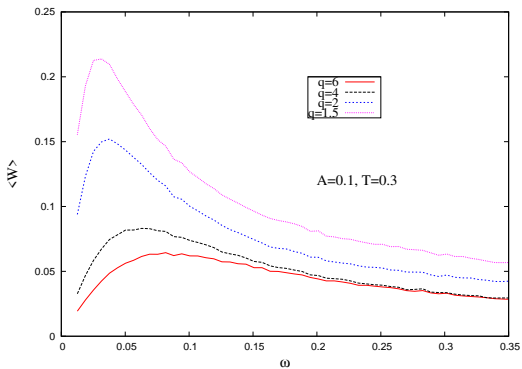}
\caption{Average work as a function of frequency for $A=0.1$ and $T=0.3$ for different values of $q$. SR is observed for all values of $q$.}
\label{SRF}
\end{figure}

\begin{figure}[h]
\includegraphics[width=7cm]{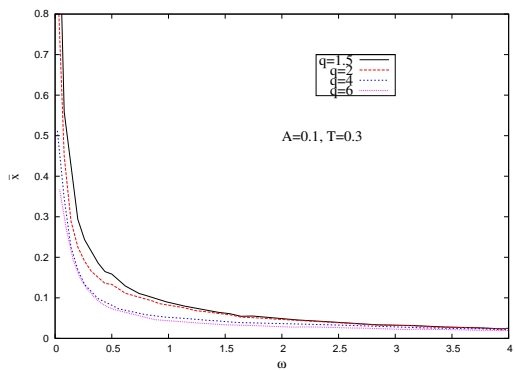}
\caption{position amplitude as a function of frequency for different values of $q$. The monotonic decrease in $\bar x$ is apparent.}
\label{xamp_om}
\end{figure}

We now study the SR quantifiers $\la W\ra$ and $\bar x$ as a function of frequency of  drive.
 The major difference between $\bar x$ and $\la W\ra$ as quantifiers is observed as a function of frequency of external drive. In figure \ref{SRF} we have plotted the mean work versus driving frequency.  Whereas $\la W\ra$ shows a peak for all values of $q$ (figure \ref{SRF}), $\bar x$ \emph{decreases} monotonically with frequency for any value of $q$ (see eq. (\ref{xTavg})), as can be seen in figure \ref{xamp_om}.
It may be emphasized here that the general trends of $\la W\ra$  and $\bar x$ as a function of $\omega$ do not depend on the nature of the confining potential. This is in contrast to the behaviour of $\la W\ra$  and $\bar x$ as a function of temperature, which crucially depends on the softness of $V(x)$.

Computing the derivative of $\la W\ra$ with respect to $\omega$ from the expression (\ref{W}), we find the maximum to occur precisely at $\omega=2r_K$:
\begin{align}
\left.\pd{\la W\ra}{\omega}\right|_{\omega_{max}} = 0 ~~\Ra ~~ \omega_{max}=2r_K.
\label{Wmax}
\end{align}
However, the proper condition for synchronization is
\begin{align}
\tau_\omega = 2\tau_K ~~\Ra ~~ \omega_{SR}=\pi r_K.
\end{align}
On calculating the escape rate for the superharmonic ($q=4$) potential at $T=0.3$, we find $r_K\approx 0.03$. Figure \ref{SRF} shows the peak to occur at $\omega\approx 0.06 \approx 2 r_K$, consistent with eq. (\ref{Wmax}).

\begin{figure}[h]
\includegraphics[width=7cm]{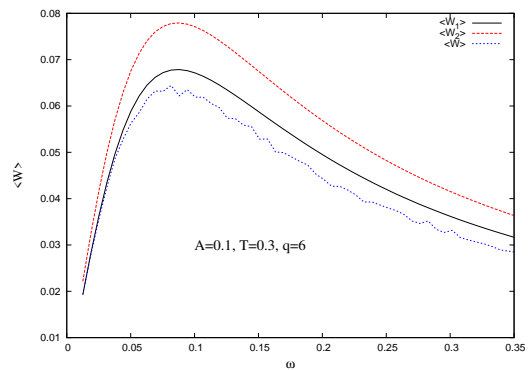}
\caption{Comparison of frequency dependence of $\la W\ra$ obtained numerically with the analytical results. $\la W_1\ra$ (the smooth solid line) is the expression obtained by taking into account the  variance in position of the particle, $\la x^2\ra_0$, at the fixed temperature $T=0.3$. $\la W_2\ra$ (the smooth dashed line) is the expression for average work obtained with the variance replaced by $x_m^2$. The curve labelled $\la W\ra$ is the numerically generated one.}
\label{compare_om}
\end{figure}

In figure \ref{compare_om}, the numerically obtained plot for $\la W\ra$ vs $\omega$ is compared with the analytical expressions (\ref{W1}) and (\ref{W2}). For this we have used the superharmonic potential with $q=6$. We observe that the analytical expression using eq. (\ref{W1}) fits better compared to (\ref{W2}), the reason being the same as explained earlier in the context of figure (\ref{ana_compare}). Now we turn our attention to the power applied to the system.

\begin{figure}[h]
\includegraphics[width=7cm]{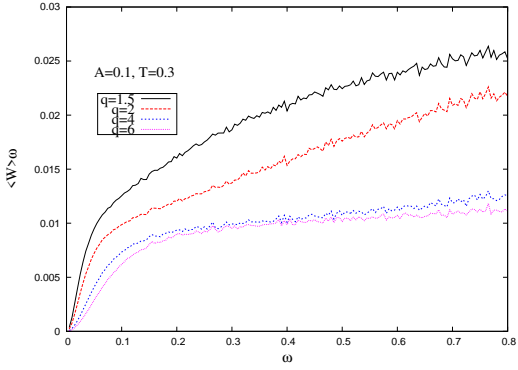}
\caption{Average power as a function of frequency for $A=0.1$ and $T=0.3$ for different values of $q$. The curves are all monotonically increasing with frequency.}
\label{power_om}
\end{figure}

In figure \ref{power_om}, we have plotted the power applied to the system by the drive versus the drive frequency. We find that for all values of $q$, the curves are monotonically increasing. From figure \ref{SRT} (scaling the y-axis by the constant parameter $\omega$), we observe that the average power exhibits peak for $q>2$ as a function of temperature. Thus, power cannot be used as a quantifier of bona fide SR \cite{jun11} for $q>2$.

\begin{figure}[h]
\subfigure[]{\includegraphics[width=7cm]{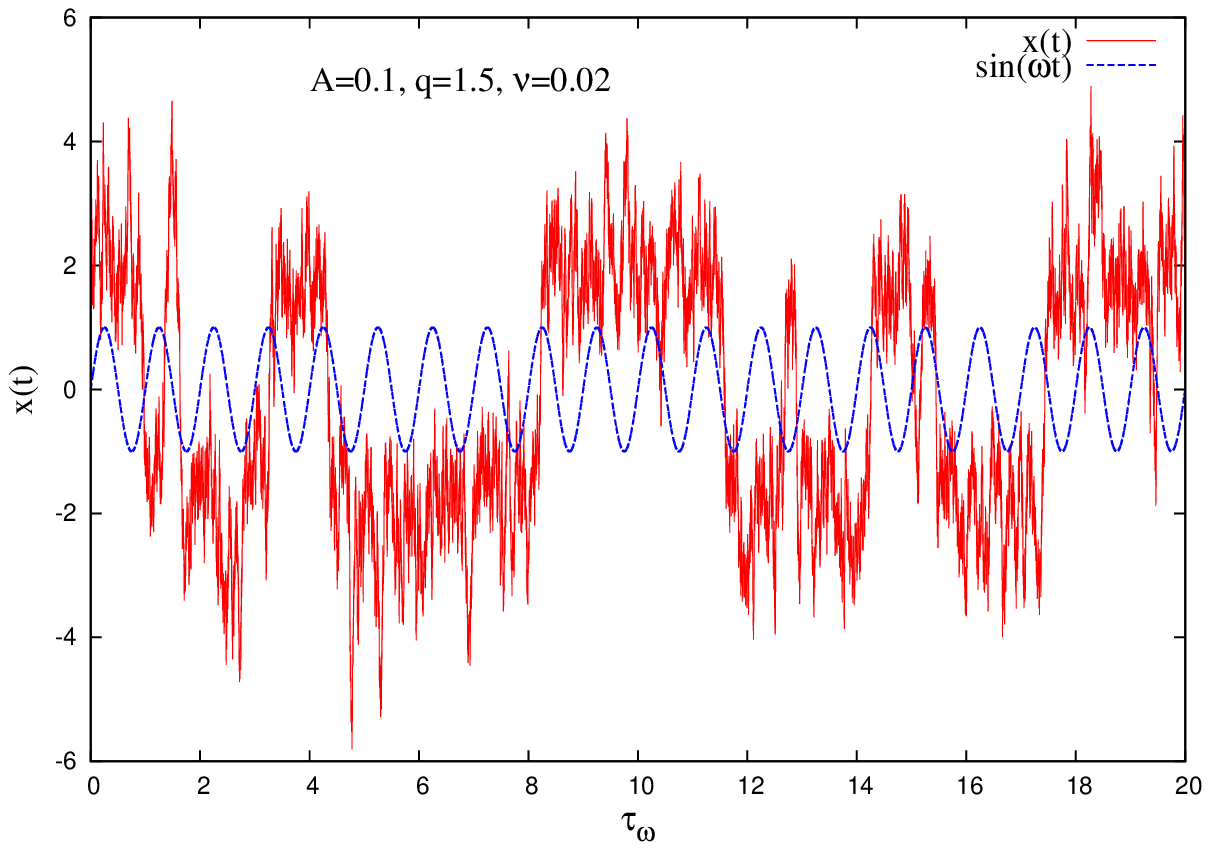}}
\subfigure[]{\includegraphics[width=7cm]{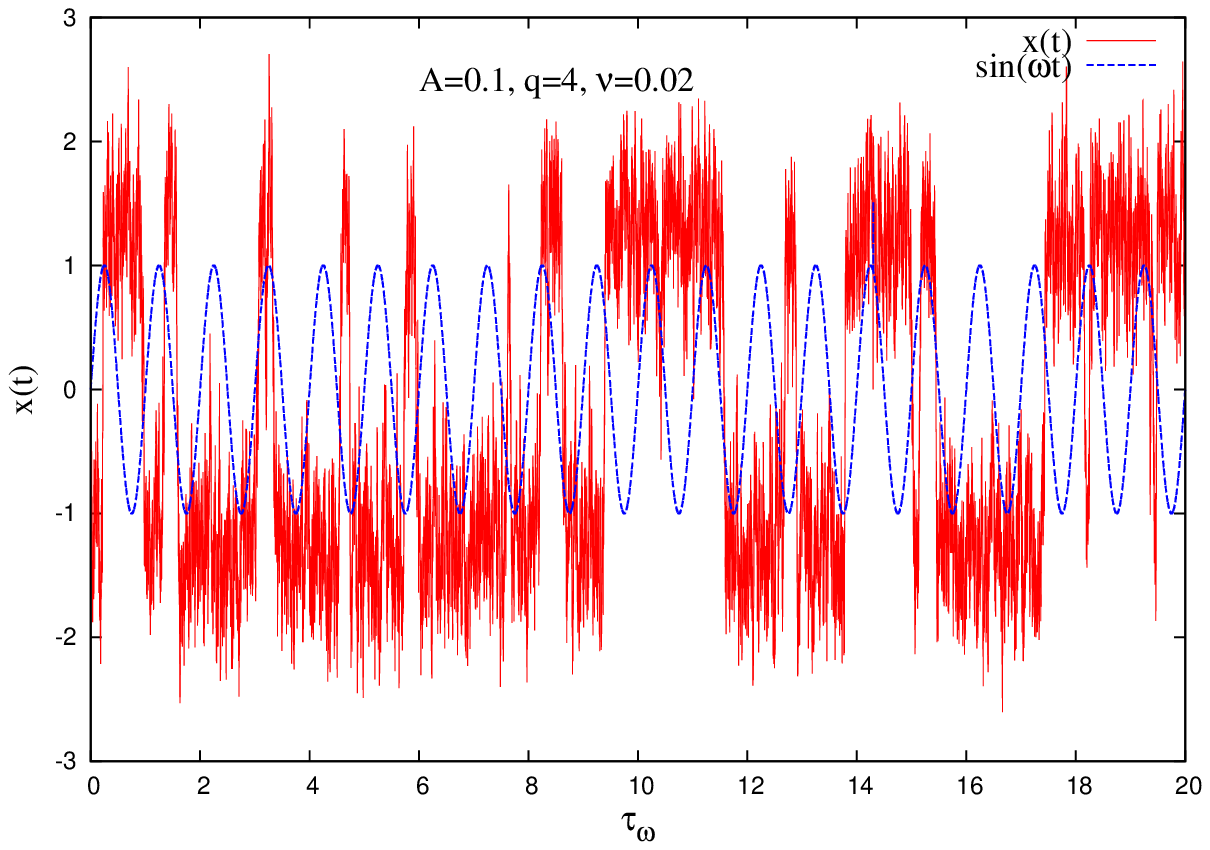}}
\caption{(a) Position as a function of time for a given trajectory is shown, for the $q=1.5$ potential. We find that the particle travels large distances from either minima due to softer confinement. (b) Similar plot for $q=4$. The hard confinement effectively contains the particle in a smaller region, and the motion is more synchronized.}
\label{realtime}
\end{figure}

\begin{figure}[h]
  \subfigure[]{\includegraphics[width=6cm]{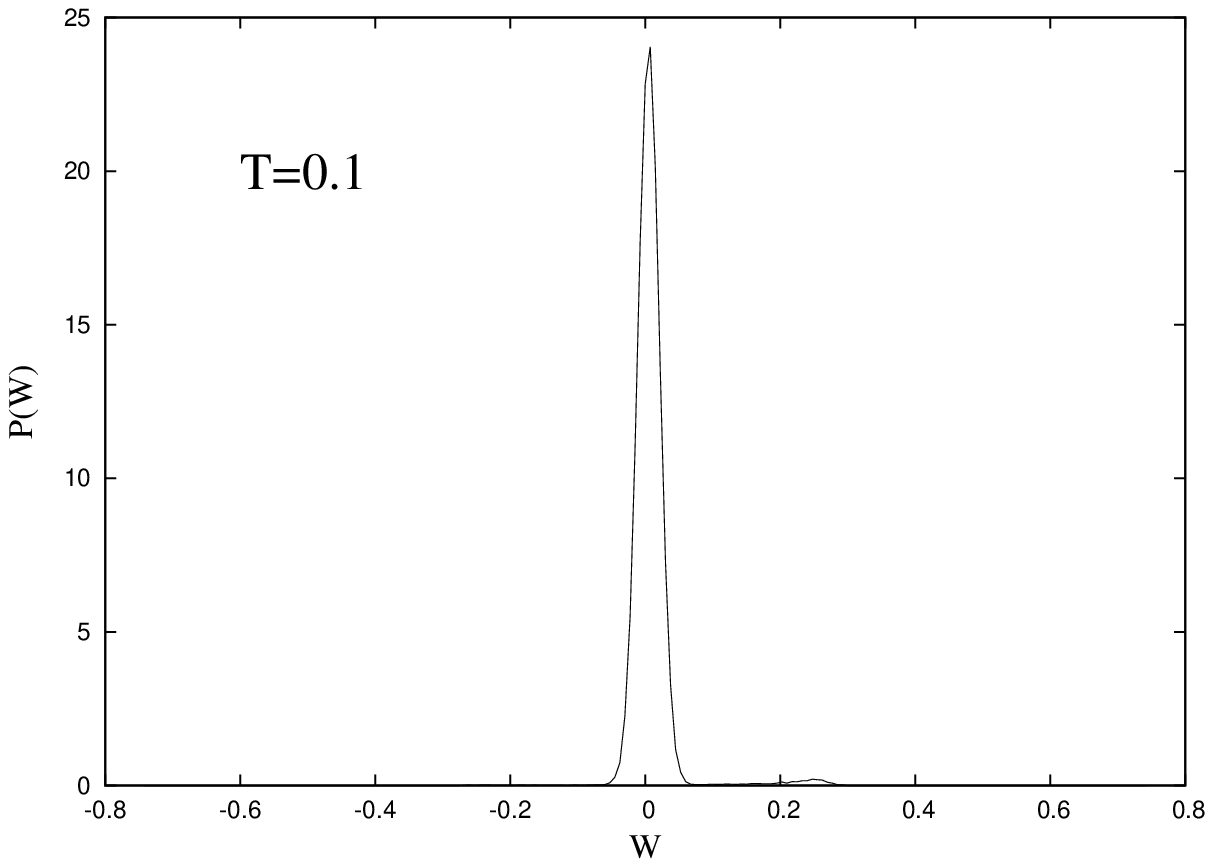}}
  \subfigure[]{\includegraphics[width=6cm]{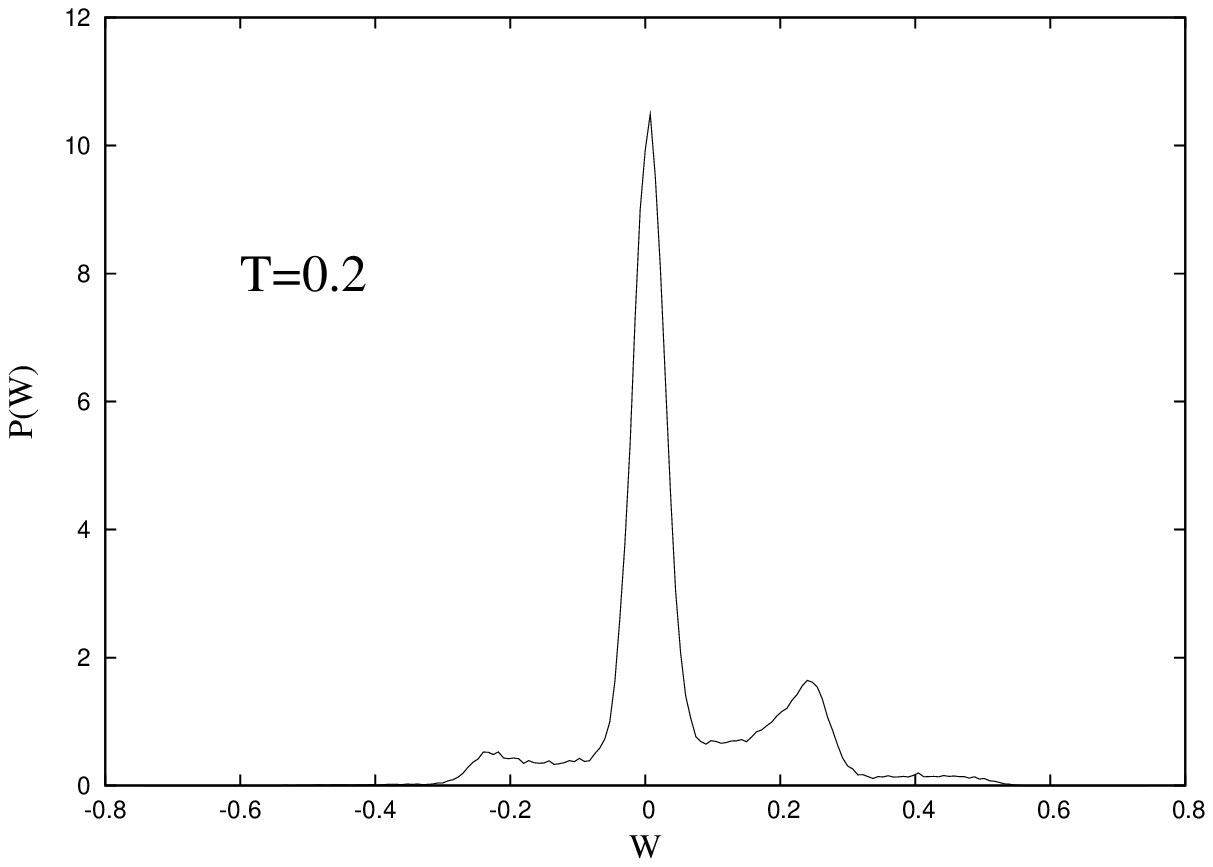}}
  \subfigure[]{\includegraphics[width=6cm]{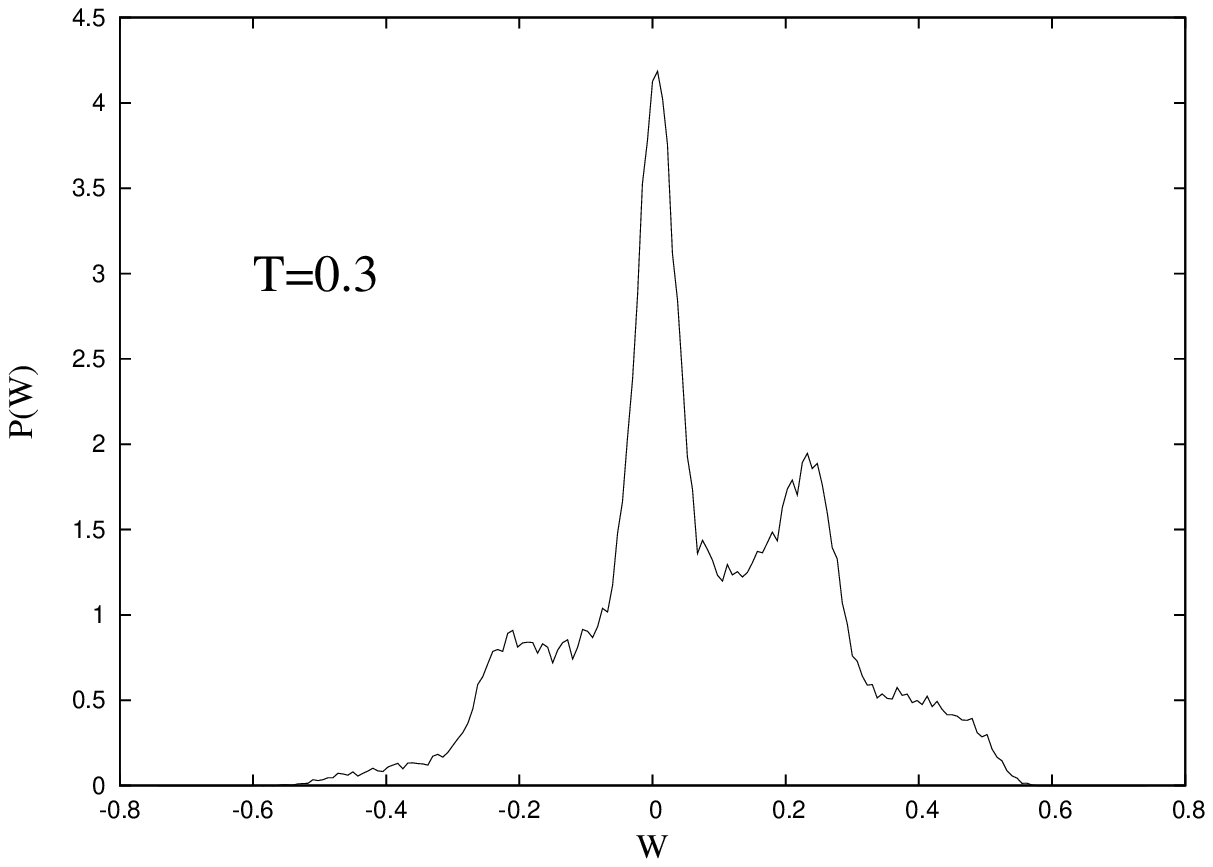}}
  \subfigure[]{\includegraphics[width=6cm]{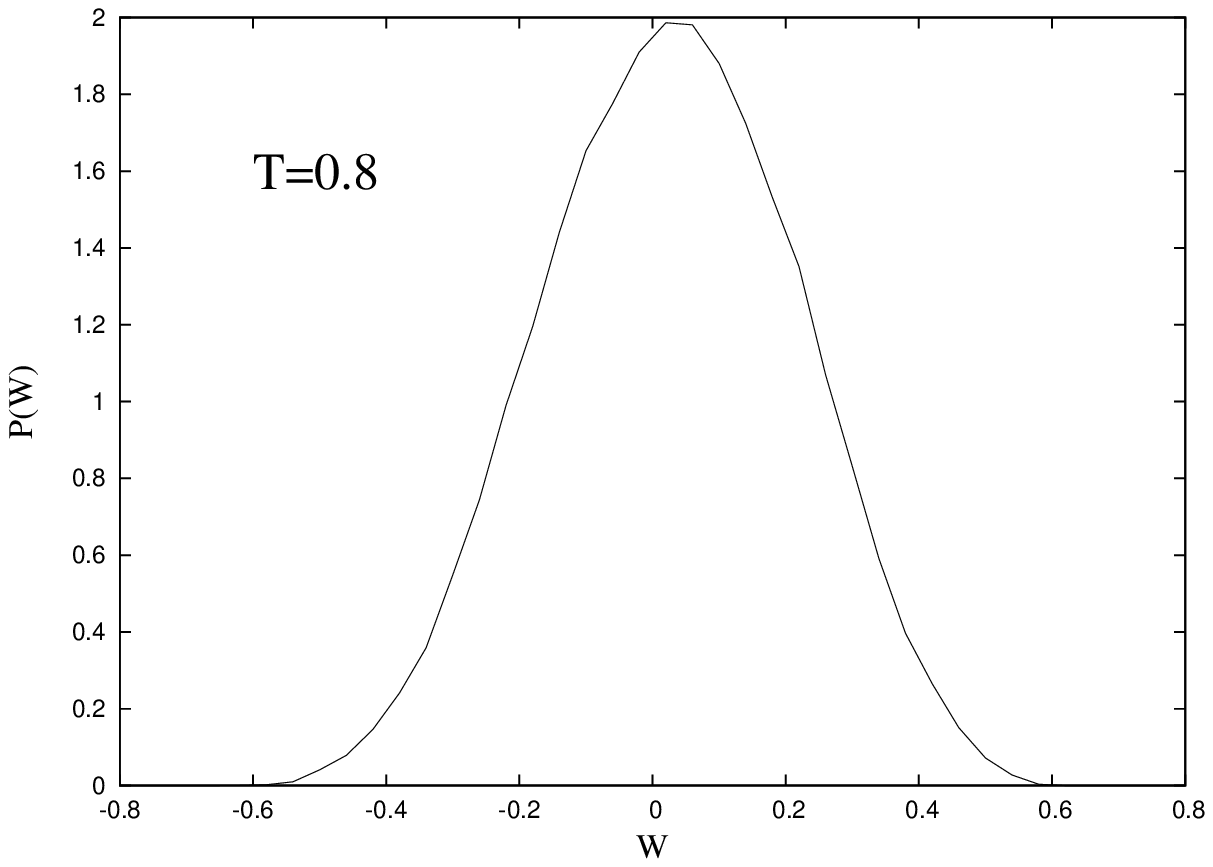}}
  \caption{Work distributions at different temperatures for q=4. Other parameters are $A=0.1$ and $\nu=0.02$.}
\label{Wdists_q4}
\end{figure}

\begin{figure}[h]
  \subfigure[]{\includegraphics[width=6cm]{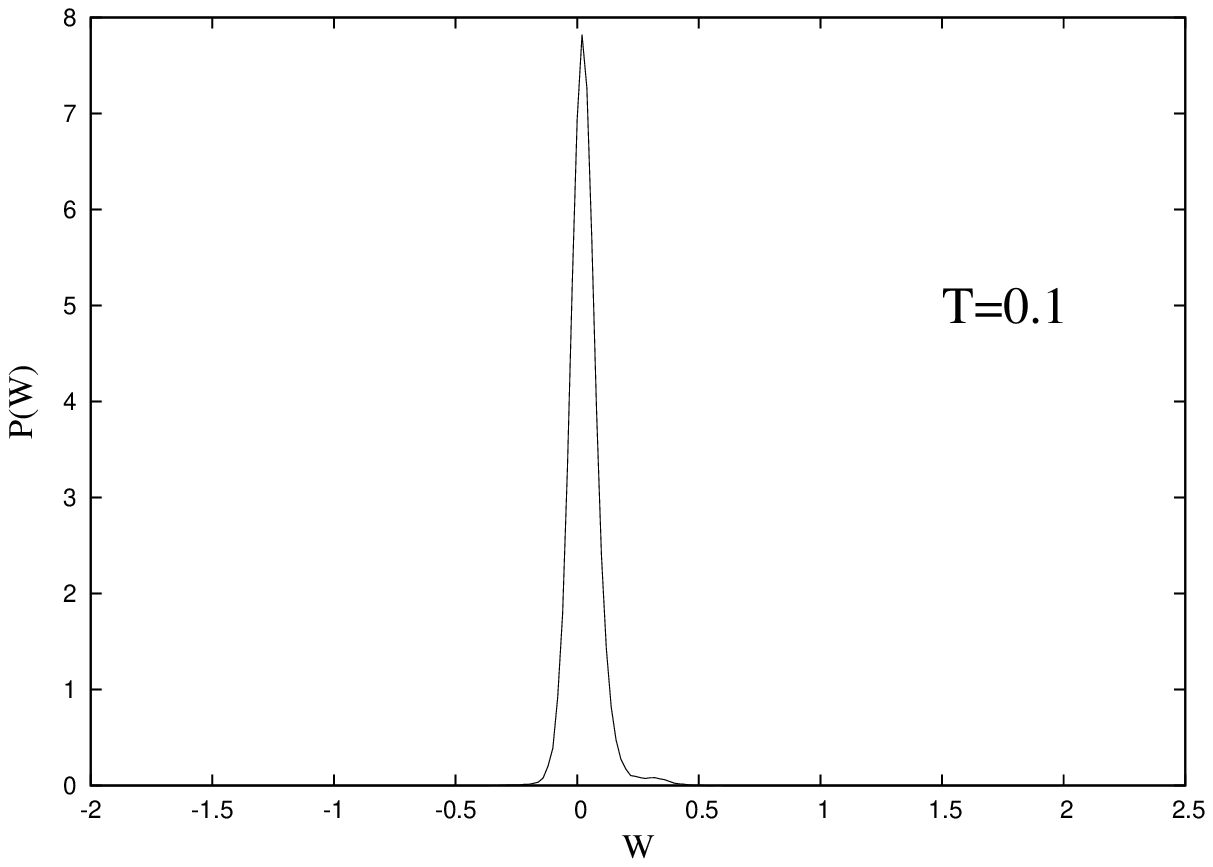}}
  \subfigure[]{\includegraphics[width=6cm]{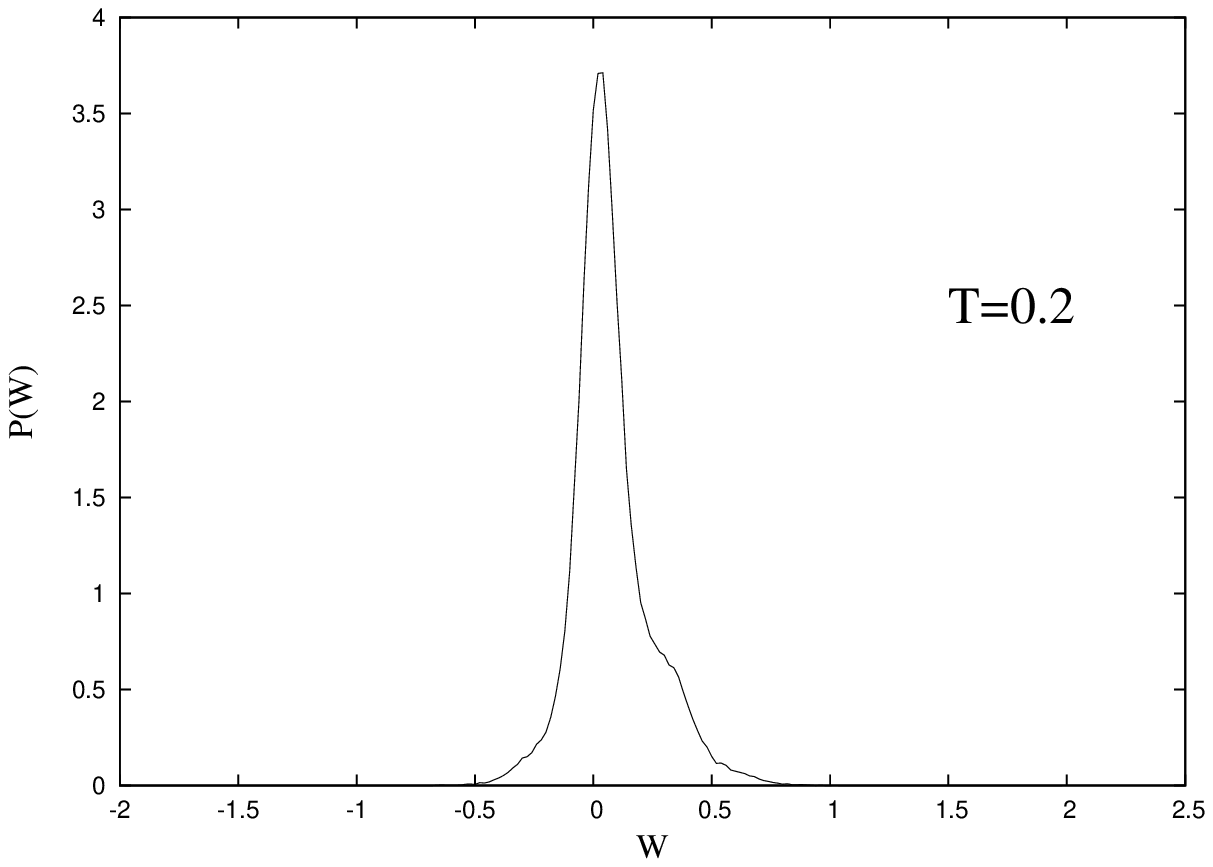}}
  \subfigure[]{\includegraphics[width=6cm]{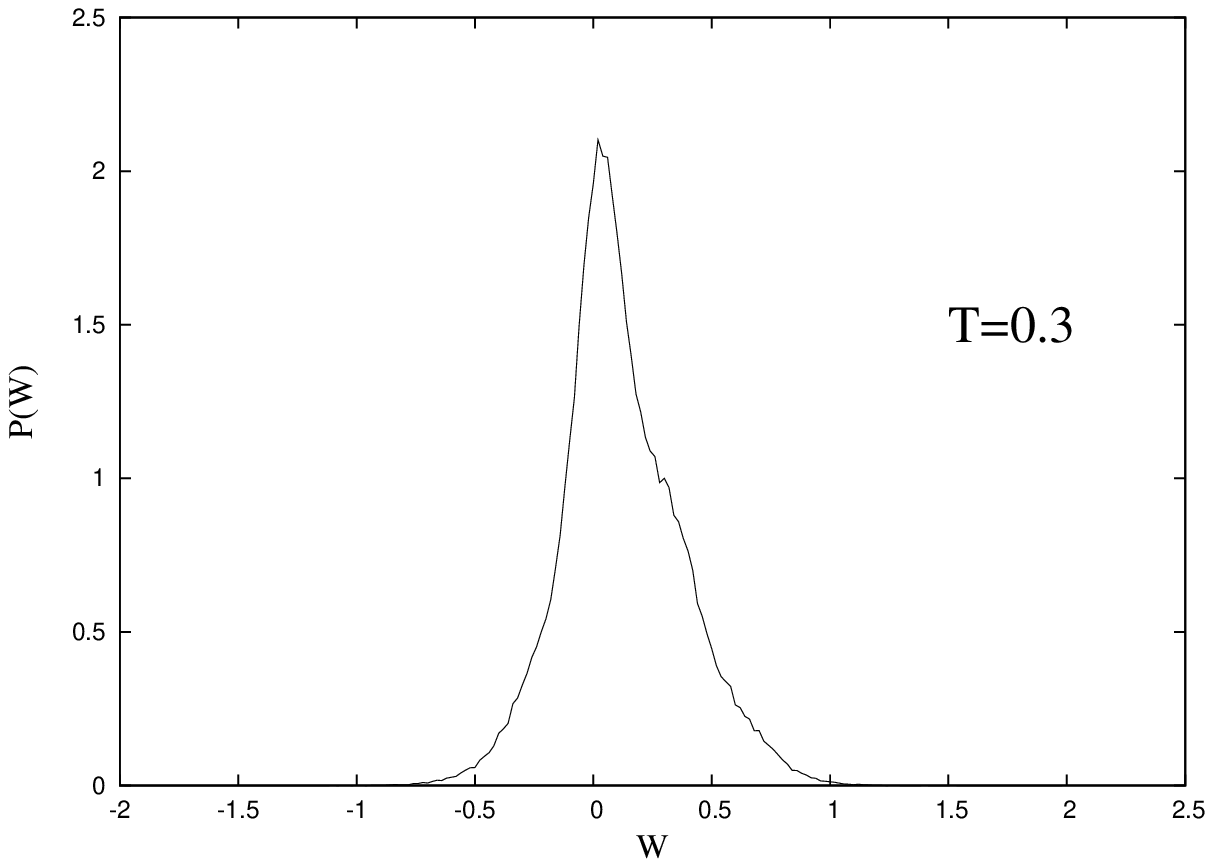}}
  \subfigure[]{\includegraphics[width=6cm]{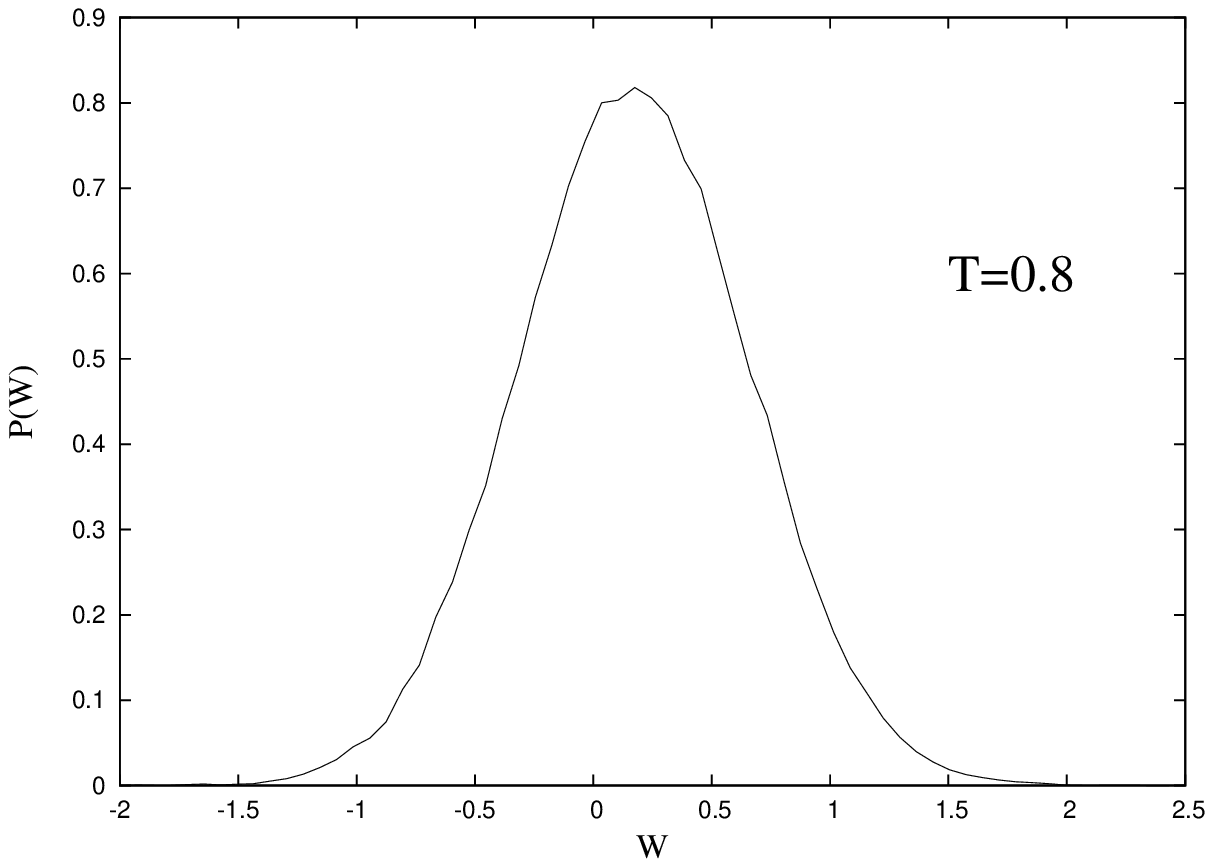}}
  \caption{Work distributions at different temperatures for q=1.5. Other parameters are $A=0.1$ and $\nu=0.02$.}
\label{Wdists_q1.5}
\end{figure}

In figure \ref{realtime}, we have plotted the real time trajectory of the particle at $T=0.3$ (around SR for q=4). In (a), because of subharmonic potential, the particle travels large distances away from the minima and spends more time in the wings of the potential ($x>x_m$ or $x<-x_m$) over a duration of many cycles of the applied force without passing over the barrier. This is clear from the figure. Thus, the question of synchronization between the applied force and particle hopping does not arise, hence the absence of SR. On the other hand, in (b), the superharmonic potential ($q=4$) helps in more efficient confinement of the particle so that the proper synchronization between the drive and the particle trajectory is attained.

In figure \ref{Wdists_q4}, we show how the work distribution $P(W)$ changes as a function of $T$ for the superharmonic potential with $q=4$. At small temperature ($T=0.1$, figure \ref{Wdists_q4} (a)), the particle sees only a single well, the barrier height being too large for it to cross. Thus the work done is entirely due to intrawell dynamics and the distribution is almost Gaussian. Occasional excursion of the particle into the other well is clearly reflected as a small hump at higher values of $W$. As $T$ increases, interwell dynamics starts playing dominant role and hence the distribution becomes broader (figures \ref{Wdists_q4} (b) and (c)). Additional peak appears towards right mainly due to the interwell motion. A third peak also appears in the negative side. For large values of temperature beyond SR point, the dynamics is dominated by interwell motion and $P(W)$ tends towards a Gaussian distribution (figure \ref{Wdists_q4} (d)).

The probability distribution for work has finite weight for negative values of $W$. These negative values correspond to the trajectories where the particle moves against the perturbing ac field over a cycle. The existence of finite weight for negative work values is essential to satisfy the recently discovered fluctuation theorems.

In figure \ref{Wdists_q1.5}, we have plotted the variations in the work distribution with temperature, for the subharmonic potential ($q=1.5$). This time, however, we do not find the appearance of prominent double and multiple peaked distribution as was observed for the superharmonic potential. This is because the particles travel higher distances during intrawell as well as interwell motion, which gives rise to work values varying over a wide range. Moreover there is no clear-cut time scale separation between intrawell and interwell motion.Thus the two distinct peaks that were observed for $q=4$ have got merged in $q=1.5$ case.

\section{Discussion and conclusions}

In this paper, we have studied the phenomenon of stochastic resonance in a bistable potential, using  the  mean input energy per cycle (or the mean work done per cycle) $\la W\ra$ as a quantifier of resonance.
We find that the system exhibits SR as a function of temperature for $q>2$, but does not show SR for subharmonic potentials. This behaviour is further verified by studying the phase lag $\bar\phi$. Thus bistability is necessary but not sufficient condition for the observation of stochastic resonance. This result is consistent with the findings in \cite{mar10}.
However, in both the superharmonic and subharmonic potentials, the work exhibits resonance peak whereas the average amplitude of mean position decreases monotonically as a function of frequency. This is quite different from the trends of $\la W\ra$ and $\bar x$ as a function of temperature, which is sensitive to the nature of the confining potential.
 We have shown that the average power delivered to the system is not a good quantifier for bonafide resonance \cite{jun11}. Our further investigation reveals qualitative differences in the nature of distributions for hard and soft potentials.

\section{Acknowledgement}

The authors thank Prof. F Marchesoni for several useful  suggestions throughout this work.
One of us (AMJ) thanks DST, India for financial support.


\begin{thebibliography}{10}

\bibitem{han98} L. Gammaitoni, P. H\"anggi, P. Jung, and F. Marchesoni, \zrmp{70}{223}{1998}.
\bibitem{mar09} L. Gammaitoni, P. H\"anggi, P. Jung and F. Marchesoni, \zepjb{69}{13}{2009}, , Special issue on stochastic resonance.
\bibitem{wie95} K. Wiesenfeld and F. Moss, Nature {\bf 373} 33 (1995).
\bibitem{wei95a} F. Moss and K. Wiesenfeld, Scientific American {\bf 273}, 66 (1995).
\bibitem{ris} H.Risken, {\it The Fokker-Planck equation}, second edition, Springer (1989).
\bibitem{cha43} S. Chandrasekhar, \zrmp{15}{1}{1943}.
\bibitem{evs05} M. Evstigneev, P. Reimann, C. Schmitt and C. Bechinger, \zjpcm{17}{S3795}{2005}.
\bibitem{iwa01} T. Iwai, \zpa{300}{350}{2001}.
\bibitem{dan03} D. Dan and A. M. Jayannavar, \zpa{345}{404}{2005}.
\bibitem{mam08} M. Sahoo, S. Saikia, M. C. Mahato and A. M. Jayannavar, \zpa{387}{6284}{2008}.
\bibitem{roy07} S. Saikia, R. Roy and A. M. Jayannavar, Physics Letters A {\bf 369}, 367 (2007).
\bibitem{mar10} E. Heinsalu, M. Patriarca and F. Marchesoni, \zepjb{69}{19}{2009}.
\bibitem{man00} R. Mannela, {\it Lecture Notes in Physics} (Springer-Verlag, Berlin, 2000) Vol. 557, p. 353.
\bibitem{jun11} P.Jung and F.Marchesoni, to be published in CHAOS, 2011.
\bibitem{dyk92} M. I. Dykman and R. Mannella, \zprl{68}{2985}{1992}.
\bibitem{sek97} Ken Sekimoto, \zptps{130}{}{1998}.

\end{thebibliography}
\end{document}